\documentclass[referee]{aa}
\usepackage{graphicx}

\begin{document}

\hyphenation{be-san-con mo-dele }

\title{Continuous optical monitoring \\ during the prompt emission of \object{GRB060111B}
\thanks{Based on observations performed with
TAROT at the Calern observatory
}
}

\author{A. Klotz\inst{1,2}
        \and B. Gendre\inst{3}
        \and G. Stratta\inst{2,4}
        \and J.L. Atteia\inst{4}
        \and M. Bo\"er\inst{2}
        \and F. Malacrino\inst{4}
        \and Y. Damerdji\inst{1,2}
        \and R. Behrend\inst{5}
}

\institute{
CESR, Observatoire Midi-Pyr\'en\'ees (CNRS-UPS), BP 4346, F--31028 -
Toulouse Cedex 04, France
\and Observatoire de Haute Provence,
    F--04870 Saint Michel l'Observatoire, France
\and IASF-Roma/INAF, via fosso del cavaliere 100, 00133 Roma, Italy
\and LATT, Observatoire Midi-Pyr\'en\'ees (CNRS-UPS), 14 Avenue E. Belin,
F--31400 - Toulouse, France
\and Observatoire de Gen\`eve, CH--1290 Sauverny, Switzerland
}

\offprints{A. Klotz, \email{klotz@cesr.fr}}

\date{Received {\today} /Accepted }

\titlerunning{Continuous optical monitoring}
\authorrunning{Klotz {\it et al.}}

\abstract
   {}
   {We present the time-resolved optical emission of \object{GRB060111B} during
   its prompt phase, measured with the TAROT robotic observatory. This is the first time
   that the optical emission from a gamma-ray burst has been continuously monitored with a
   temporal resolution of a few seconds during the prompt gamma-ray phase.}
   {The temporal evolution of the prompt optical emission at the level of several seconds
   is used to provide a clue to the origin of this emission.}
   {The optical emission was found to decay steadily from our first
    measure, 28s after the trigger, in contrast to
   the gamma-ray emission, which exhibits strong variability
   at the same time. This behaviour strongly suggests that the optical emission
   is due to the reverse shock.}
   {}

\keywords{gamma-ray : bursts }

\maketitle

\section{Introduction}

\object{GRB060111B} was a bright, double--peak, gamma-ray burst (GRB) detected on January 11$^{th}$, 2006,
at 20:15:43.24 UT (hereafter t$_\mathrm{trig}$) by the BAT instrument on the Swift spacecraft
(trigger=176918, Perri et al.~\cite{Perri2006}, Tueller et al.~\cite{Tueller2006}).
The first peak started at t-t$_\mathrm{trig}$=-4 sec, peaked at t-t$_\mathrm{trig}$=+1,
and extended out to t-t$_\mathrm{trig}$=+28 sec.
The second, smaller peak started at t-t$_\mathrm{trig}$=+53, peaked at t-t$_\mathrm{trig}$=+55,
and was over by t-t$_\mathrm{trig}$=62 sec.
The duration T$_\mathrm{90}$ was 59 $\pm$1 sec in the range 15-350 keV.
The power--law index of the time-averaged spectrum (t$_\mathrm{trig}$-2 to t$_\mathrm{trig}$+63 sec)
was 1.04$\pm$0.17.  The fluence in the 15-150 keV band was
(1.6$\pm$0.1)$\times$10$^{-6}$ erg/cm$^2$.  The 1s peak photon flux measured
from t$_\mathrm{trig}$+0.36 sec in the 15-150 keV band was 1.4$\pm$0.3 ph/cm$^2$/sec.
No optical redshift was available when this paper was written.
Perri et al.~(\cite{Perri2006}) mentions that UVOT detected the
afterglow in a filtered B image taken about five minutes after the trigger.
If we conservatively place the Lyman alpha limit at the end of
the B bandpass (480\,nm, Bessel~\cite{Bessel1990}), \object{GRB060111B} is at $z$$<$3.0.

In this letter we report early optical observations
of \object{GRB060111B},
performed with the TAROT robotic observatory.
The gamma-ray burst coordinates network (GCN) notice, providing celestial
coordinates to ground stations, was sent at 20:16:03 (Perri et al.~\cite{Perri2006}).
The first image of TAROT started at 20:16:11.22 UT, 27.98s after the
GRB (9s after the notice) and 34s before the end of the high--energy emission.
From our images, we derived its position: R.A. =
19h 05m 42.46s and Dec.= +70$^{\circ}22'32.3''$ (J2000.0) with an
accuracy of 1 arcsec.
The observations presented here provide the first opportunity
to check the rapid variability
of the optical emission during the prompt phase of a gamma-ray burst.

\section{TAROT observations}
\label{obsdata}

\begin{figure}[htb]
\centering{\includegraphics[width=0.8\columnwidth]{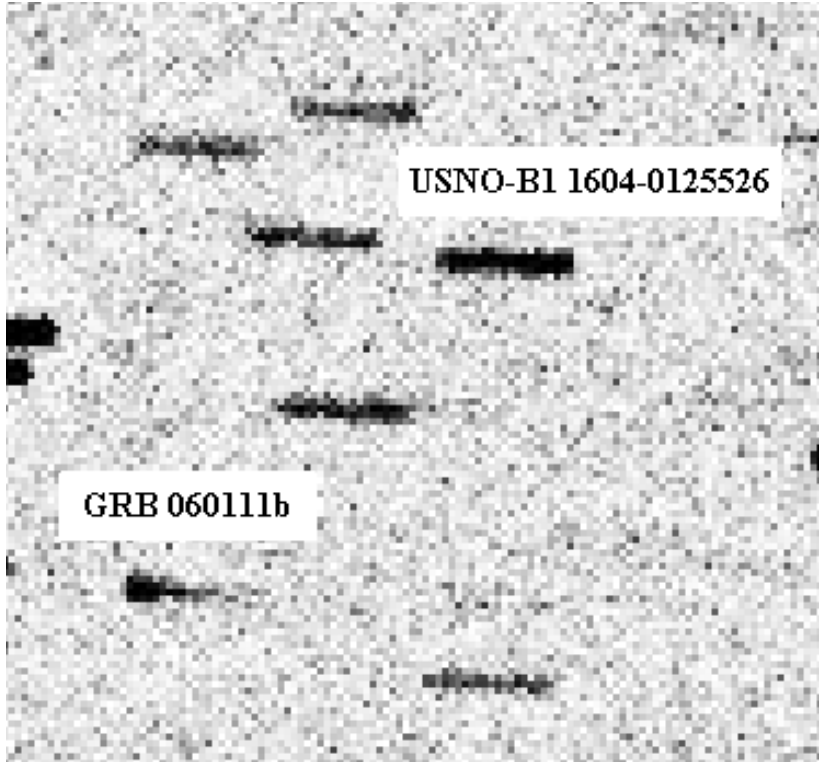}}
\caption{
This image was taken between 28s and 88s after the GRB trigger.
The hour angle speed was adapted to obtain stars as trails
of $\sim$18 pixel length during the 60s exposure. The light--curve
was then time--resolved with no dead time. In this image, the nearby star
USNO-B1 1603-0125834, close to the GRB, was removed using the trail
of USNO-B1 1604-0125526.
} 
\label{trail}
\end{figure}

TAROT is a fully autonomous 25 cm aperture
telescope installed at the Calern observatory (Observatoire de la
Cote d'Azur - France)
devoted to very early observations of GRB~optical counterparts.
A technical description can be read in Klotz et al.~(\cite{Klotz2005}).
The first image of \object{GRB060111B} is a 60s exposure taken in drift mode.
The tracking of the hour--angle motor was adapted to
a drift of 0.30 pixels/s.
As a consequence, stars are trailed with a length of about
18 pixels on the image (see Fig.~\ref{trail})
and the flux was recorded continuously with no dead time.
Subsequent images were recorded as a series of
30\,s exposures tracked on the diurnal motion.
All images were taken with no filter.
The on--line pre--processing software provided calibrated images
(dark--corrected, flat--fielded and astrometrically calibrated
from the USNO-A2.0 catalog).
The optical transient is detected in the first 4 images.
Subsequent images were co-added to increase the signal-to-noise ratio
but the afterglow was not detected.


From the trailed image, the optical light--curve 
is directly obtained by a binning in the declination direction.
We chose USNO-B1 1604-0125526 (R=13.90) as the
reference star for flux calibration.
Stars USNO-B1 1603-0125834 (R=15.79) and
1603-0125830 (R=16.40) were close to the optical transient,
and were removed by an appropriate scaling of the reference star.
From t$_\mathrm{trig}$+28s to t$_\mathrm{trig}$+61s,
we used a time sampling of one pixel (3.33s).
After 61s we used a four--pixel time sampling (13.3s)
until the end of the trail to increase the signal--to--noise ratio.
For the non--trailed images, we fit the optical transient
by the PSF of the reference star to deduce magnitudes.
We designated as CR the magnitudes computed from our
unfiltered images calibrated with the reference star.
Differences between CR and R magnitudes come from uncertainties
in the calibration of the R magnitude of the reference star
and from colour differences with the optical transient.
Table \ref{logobstable} gives the results. Due to the
unknown accuracy of the USNO-B1 catalog,
a global shift of $\pm$0.2 magnitude cannot be excluded.

\begin{table}[htb]
\caption{Log of the optical measurements from TAROT. T are
seconds since $t_\mathrm{trig}$. Uncertainties on CR magnitudes
are 2$\sigma$ errors.}
\begin{center}
{\scriptsize
\begin{tabular}{c c c | c c c}
T$_\mathrm{start}$ & T$_\mathrm{end}$ & CR & T$_\mathrm{start}$ & T$_\mathrm{end}$ & CR \cr
\noalign{\smallskip} \hline \noalign{\smallskip}
 28.0  &  31.3  & 13.75   $\pm$ 0.12    &  54.6  &  58.0  & 15.58   $\pm$ 0.35    \cr
 31.3  &  34.6  & 13.90   $\pm$ 0.19    &  58.0  &  61.3  & 15.40   $\pm$ 0.45    \cr
 34.6  &  38.0  & 14.05   $\pm$ 0.18    &  61.3  &  74.6  & 15.79   $\pm$ 0.19    \cr
 38.0  &  41.3  & 14.43   $\pm$ 0.27    &  74.6  &  87.9  & 16.31   $\pm$ 0.27    \cr
 41.3  &  44.6  & 14.63   $\pm$ 0.26    &  94.6  & 124.6  & 16.46   $\pm$ 0.30    \cr
 44.6  &  48.0  & 14.94   $\pm$ 0.24    & 131.5  & 161.5  & 16.89   $\pm$ 0.38    \cr
 48.0  &  51.3  & 15.11   $\pm$ 0.40    & 168.3  & 198.3  & 17.53   $\pm$ 0.59    \cr
 51.3  &  54.6  & 15.30   $\pm$ 0.28    &        &        &                       \cr
\noalign{\smallskip} \hline
\end{tabular}
}
\label{logobstable}
\end{center}
\end{table}



\section{The prompt emission and early afterglow}
\label{grb060111B}

\begin{figure}[htb]
\includegraphics[width=\columnwidth]{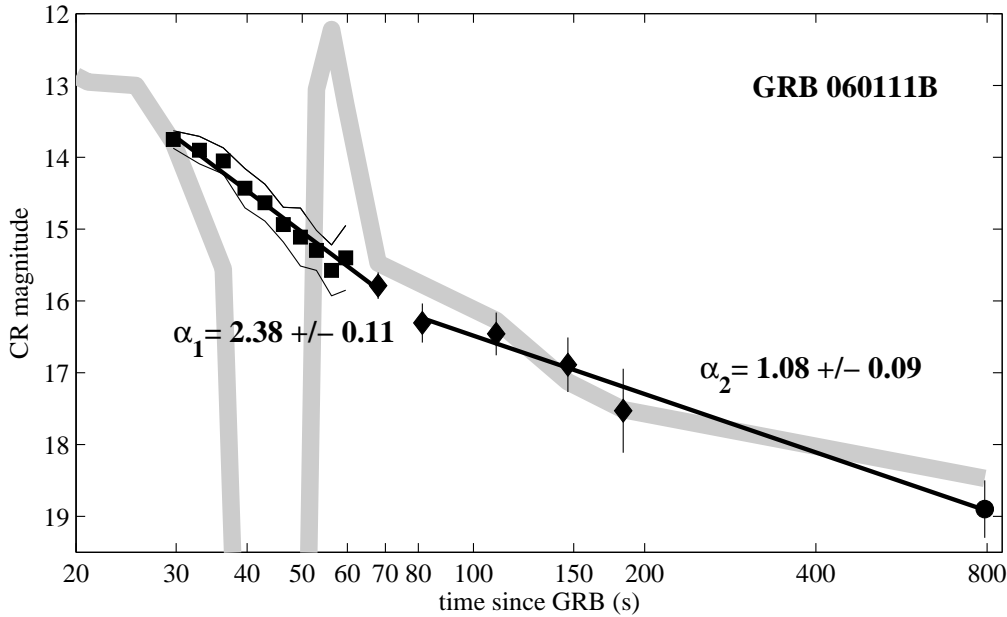}
\caption{
light--curves of \object{GRB060111B}.
Black squares and diamonds with vertical error bars are
CR magnitudes measured by TAROT (see Table \ref{logobstable}).
The black circle indicates the R magnitude measured by MITSuME
(Yanagisawa et al.~\cite{Yanagisawa2006}).
The solid straight lines are fits by a $f(t) \propto t ^{-\alpha}$ law.
The broad grey line shows the high--energy light--curve
($-2.5\,\log$(X-ray flux)) vertically offset to
be compared to the optical light--curve.
Before 75s we extrapolated Swift-BAT flux to the 2-10~keV
energy range.
After 90s, we used the XRT data in the 2-10~keV energy range
(see Table~\ref{tab2}).
} \label{lc}
\end{figure}

The optical light--curve of \object{GRB060111B} 
is presented in Fig. \ref{lc}. The optical
emission exhibits two phases. A fast decaying phase
with a slope $\alpha_1$ = 2.38$\pm$0.11 dominates
from 28s to 80s after the trigger,
roughly corresponding to the first TAROT exposure with high temporal resolution.
This fast decay was also observed by ROTSE-IIId 
(Yost et al.~\cite{Yost2006}).
This phase is followed by a period of shallower decay with a slope
$\alpha_2$ = 1.08$\pm$0.09, typical of the late afterglow phase.
The observation
of  Yanagisawa et al.~(\cite{Yanagisawa2006}) in the R band
at t$_\mathrm{trig}$+800s
is compatible with the extrapolation of the TAROT points
with this slope.
The transition between the two phases occurred at
t$_\mathrm{trig}$+80s, about 20s after the end of the gamma-ray emission.
These slopes and transition time were derived under the assumption
that there is a single
emission component evolving from a steep to a shallow decay.
If we assume instead that the optical emission is due to the
superposition of two components that are simultaneously
present (e.g. the forward and the reverse shocks),
the slopes of the fast and shallow decays become
$\alpha_1$=3.0 and $\alpha_2$=0.89, respectively, and the transition occurs
at t$_\mathrm{trig}$+65s, right at the end of the gamma-ray
emission.

A remarkable feature of the prompt optical light--curve is its smoothness.
As can be seen in Fig.~\ref{lc}, the broken power--law fit
adequately reproduces the evolution of
the optical emission from t$_\mathrm{trig}$+28s to t$_\mathrm{trig}$+200s.
All individual measurements are consistent with a smooth decay
of the optical emission.
This is the first time
that the optical variability of a gamma-ray burst
could be measured during the prompt phase,\footnote{We call 'prompt phase' the
period during which SWIFT/BAT detected the high--energy emission. This period
extended from 2s before the trigger to 62s after the trigger.}
down to timescales of $\sim$5s.
Figure~\ref{lc} also displays the high--energy light--curve in the 2-10 keV
energy range (the broad grey line). Before 75s, the 2-10 keV flux was extrapolated from the
flux measured with the BAT in the 15-150 keV energy range, assuming a spectrum
with a photon index $\Gamma$=-1.57, in agreement with the prescription
of O'Brien et al.~(\cite{Obrien2006}) recommending use of a photon
index that is the average of the values measured in the
BAT and in the XRT.
After 90s, the 2-10 keV light--curve was measured by the XRT.
The high--energy (2-10 keV) fluxes in time intervals that are
simultaneous with the optical intervals are given in Table~\ref{tab2}.
The high--energy emission is highly variable, with two well--defined peaks.
The variability of the high--energy emission contrasts with the
smoothness of the optical emission observed at the same time.
In particular, we note that there is no significant variation
in the optical flux at the time of the second gamma-ray peak
at t$_\mathrm{trig}$+55s.
After t$_\mathrm{trig}$+90s, the XRT light--curve can be described by
a simple power--law with slope 1.1$\pm$0.1, which is fully compatible
with the slope of the optical emission.

\begin{figure}[htb]
\centering{\includegraphics[width=0.85\columnwidth]{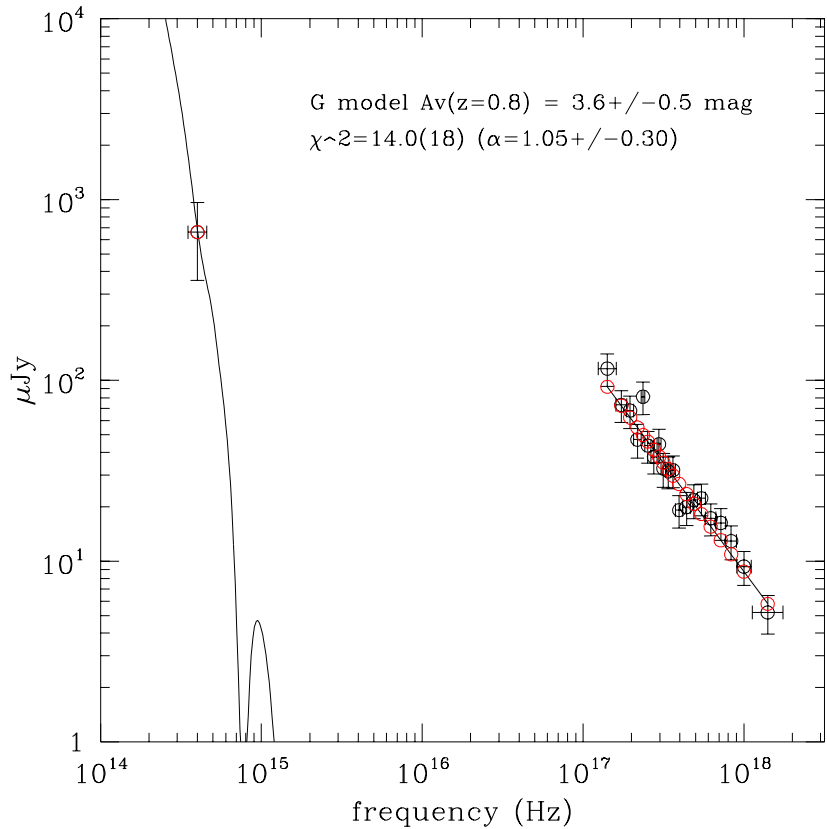}}
\caption{The optical-to-X-ray afterglow spectrum from 95s to 198s after
the burst. The X-ray data have been corrected for both the
Galactic and the extra-galactic absorption. The optical data have been
corrected only from the Galactic extinction. The solid line is the 
best--fit absorbed power--law model. We assumed a local extinction curve
similar to the one observed in our Galaxy. The spectral index was fixed
within the errors to the value derived only from the X-ray analysis.
}
\label{spectrum}
\end{figure}

Figure~\ref{spectrum} shows the broad-band spectrum of the emission
during the early afterglow from t$_\mathrm{trig}$+95s to t$_\mathrm{trig}$+200s.
The energy index of the 0.5-10 keV spectrum is 1.05 $\pm$ 0.3,
a typical value for an afterglow.
The X-ray spectrum requires an absorption corresponding to
N$_\mathrm{H}$= 2.8$\pm$0.8 10$^{21}$ cm$^{-2}$,
significantly exceeding the galactic value, N$_\mathrm{H}$= 0.7 10$^{21}$ cm$^{-2}$.
The extrapolation of the unabsorbed X-ray spectrum to the optical range overpredicts
the optical emission, indicating the necessity of a break between
the optical and X-ray domains.
This break could reflect the spectral distribution of the emission or
a strong optical extinction.
With the data at hand, it is difficult to choose between these
two possibilities. Nevertheless if we consider the consistency of the decay rate
at optical and X-ray wavelengths as an indication of a common origin,
we may attribute the observed spectral break to optical extinction.
An extinction of A$_\mathrm{R}$=4 magnitudes is required to
reconcile the measured optical flux with the extrapolation of the
XRT spectrum.
We note that a dense environment could explain many features of \object{GRB060111B}:
a reverse shock peaking early (see next section), a large column density
of hydrogen, and a possible large optical extinction.

Assuming a standard galactic extinction law, the optical 
extinction is compatible with
the hydrogen column density measured in X-rays (N$_\mathrm{H}$), if the source is situated
at z $\sim$ 0.8 (and A$_\mathrm{v}$=3.6 at the source).
This is the situation illustrated in Fig.~\ref{spectrum}.
Given the errors on the measurement of N$_\mathrm{H}$, the consistency
between the expected and the measured N$_\mathrm{H}$/Av ratio is
maintained up to z$_\mathrm{max}$=1.2.
Larger redshifts (up to $z$=2.5) are also possible if we assume
a non-standard extinction law with a weaker wavelength dependence than
for the Galactic case (Stratta et al.~\cite{Stratta2004}).

\begin{table}[htb]
\caption{2-10 keV light--curve of GRB 060111B.
Columns 1,2,4, and 5 give the start and end times
of the intervals in seconds after the trigger.
Columns 3 and 6 give the high--energy fluxes and their errors
in units of 10$^{-10}$ erg cm$^{-2}$ s$^{-1}$.
Fluxes before t$_\mathrm{trig}$+75 sec are based on an extrapolation
of the BAT counts to the 2-10 keV range (see text). Fluxes after 90 sec are
measured by the XRT. All fluxes have been corrected for absorption
considering a column density of hydrogen of N$_\mathrm{H}$= 2.8 10$^{21}$ cm$^{-2}$}
\begin{center}
{\scriptsize
\begin{tabular}{c c c | c c c}
T$_\mathrm{start}$ & T$_\mathrm{end}$ & $flux$ & T$_\mathrm{start}$ & T$_\mathrm{end}$ & $flux$ \cr
\noalign{\smallskip} \hline \noalign{\smallskip}
         0.1    &     0.50  &  190 $\pm$  37  &       25.3    &    27.6  &   45 $\pm$  10  \cr
        0.50    &     1.13  &  260 $\pm$  40  &       27.6    &    33.0  &   22 $\pm$  11  \cr
        1.13    &      2.0  &  210 $\pm$  27  &       33.0    &    41.3  &  4.3 $\pm$  5.3 \cr
         2.0    &      3.0  &  180 $\pm$  27  &       41.3    &    49.6  &    0 $\pm$  4.1 \cr
         3.0    &      4.0  &  120 $\pm$  25  &       49.6    &    54.6  &   43 $\pm$  6.8 \cr
         4.0    &     5.25  &   88 $\pm$  25  &       54.6    &    58.0  &   92 $\pm$  7.0 \cr
        5.25    &        7  &   88 $\pm$  18  &       58.0    &    63.8  &   35 $\pm$  6.5 \cr
           7    &        9  &   80 $\pm$  18  &       63.8    &    74.6  &  4.6 $\pm$  3.2 \cr
           9    &     11.5  &   47 $\pm$  18  &       94.6    &   124.6  &  2.2 $\pm$  0.3 \cr
        11.5    &       15  &   84 $\pm$  12  &      131.5    &   161.5  &  1.0 $\pm$  0.3 \cr
          15    &       19  &   66 $\pm$  12  &      168.3    &   198.3  &  0.7 $\pm$  0.2 \cr
          19    &     23.3  &   47 $\pm$  12  &      762.0    &   822.0  & 0.29 $\pm$ 0.09 \cr
\noalign{\smallskip} \hline
\end{tabular}
}
\label{tab2}
\end{center}
\end{table}

\section{Interpretation of the observations}
\label{comparison}

The nature of the optical light contemporaneous with the prompt
gamma-ray emission remains an important, yet open, issue.
In the internal/external shock model, the 
prompt optical emission could be due to internal shocks within the
relativistic ejecta and/or to the reverse shock crossing the ejecta. 
In both cases the optical radiation
carries crucial information on the composition of the ejecta. 
This information is lost at later times, when the optical emission 
is dominated by the afterglow from the external shock. 
The first step on the (long) way to the comprehension of GRB ejecta 
is thus to clarify whether the optical emission originates in internal
shocks or in the reverse shock. 
This issue has been discussed for a few GRBs in which the prompt 
optical emission could be detected, but with diverse conclusions.

The bright optical flare of \object{GRB990123} (Akerlof et al.~\cite{Akerlof1999})
has been attributed 
to reverse--shock emission (Sari \& Piran \cite{Sari1999a},
M{\'e}sz{\'a}ros \& Rees \cite{Meszaros1999},
Wang et al. \cite{Wang2000} and ref. therein).
This interpretation had important implications for the physical
properties and composition of the ejecta and of the surrounding
medium (e.g. Sari \& Piran \cite{Sari1999b}, 
Fan et al. \cite{Fan2002}, Zhang et al. \cite{Zhang2003}). 
On the other hand, 
Vestrand et al. (\cite{Vestrand2005}) find that
the optical emission of \object{GRB041219A} is proportional to the
high--energy emission of this burst, which is the behaviour 
expected if the optical emission 
is due to internal shocks, although this conclusion
is based on three points with a positive detection,
which might also be explained by a bright optical
flare followed by a steep decay.
The case of \object{GRB050904} is less clear because the optical
flare is almost simultaneous with a strong X-ray flare
attributed to the prompt emission (Bo{\"e}r et al.~\cite{Boer2006},
Wei et al.~\cite{Wei2006}).


The TAROT observations presented here allow a first 
investigation of the details of the temporal structure of the
prompt optical emission, thereby providing crucial data to resolve the
question of the origin of this emission.
Such a study has not been possible till now,
due to the shortness of the bursts (\object{GRB021211}, \object{GRB050319},
\object{GRB050401}, \object{GRB050801}),
to a low temporal resolution (\object{GRB990123}), or to the
faintness of the optical emission (\object{GRB041219A}
\& \object{GRB050904}).
Our observations reveal the lack of fine time structure
during the decay of the bright optical flare of \object{GRB060111B}.
This result is particularly significant since, during the same time
interval, the gamma-ray light--curve
shows a peak not seen in optical observations.
The lack of correlation between the optical emission and the
high--energy radiation eliminates internal shocks as the main 
origin of the optical light seen by TAROT. In consequence, the single flash
detected by TAROT and ROTSE-IIId must be attributed to the reverse shock
or to a forward shock peaking early. The fast decay 
of the optical emission with $\alpha_1$ = 2.4, strongly suggests that it comes
from the reverse shock. 

The peak of the forward
shock is not detected and must be below the level of the reverse shock emission; this is
the Type II light--curve studied by Zhang et al. (\cite{Zhang2003}).
It is generally assumed that the physical parameters of the reverse shock 
and of the forward shock are similar (e.g., Sari \& Piran 1999a, b;
M\'esz\'aros \& Rees 1999; Wang et al. 2000; Wei 2003). 
In that case, however, as shown in 
Zhang et al. (2003) and Fan et al. (2005), the reverse shock 
optical emission can not dominate over the forward shock optical emission when the
latter peaks. It therefore looks surprising that no forward shock peak is  
evident in the optical lightcurve. Notwithstanding, such a situation has been observed
in \object{GRB 990123}, \object{GRB 021211}, and the current burst \object{GRB 060111B}
and may be common among
GRBs. Such light--curves suggest that the physical parameters in the reverse shock are
much larger than in the forward shock, in particular, the fraction of 
shock energy given to the magnetic field. 
The thermal energy density contained in both shocks being comparable,
the magnetic field in the reverse shock is thus much stronger than the one in the
forward shock. Type II light--curves have been used as an indication that
the GRB outflow is magnetized (e.g., Fan et al. 2002; Zhang et al. 2003).
Detailed simulations
of the pan-chromatic emission of \object{GRB060111B}, including the prompt
emission, the unambiguously identified reverse shock, and the early forward 
shock are beyond the scope of this paper; yet it can be expected that such
simulations will bring new insight into the nature of GRB ejecta.



%
\begin{acknowledgements}
B. G. and G. S. acknowledge the support by the EU Research and
Training Network "Gamma-Ray Bursts, an Enigma and a Tool."
The TAROT telescope is funded by the {\it Centre National
de la Recherche Scientifique} (CNRS), {\it Institut National des
Sciences de l'Univers} (INSU), and the Carlsberg Fundation. It was
built with the support of the {\it Division Technique} of
INSU. We thank the technical staff contributing to the TAROT project:
G. Buchholtz, J. Eysseric, C. Pollas and Y. Richaud. The authors thank
the anonymous referee for very helpful comments on the interpretation 
of the observations.

\end{acknowledgements}

\end{document}